\documentclass{rspublic}
\usepackage[utf8]{inputenc}
\usepackage{graphicx,t1enc,color}
\usepackage[square,comma]{natbib}

\newcommand{\degree}{\mbox{\r{}}}

\newcommand{\eqnref}[1]{Eq.~\ref{eq:#1}}
\newcommand{\picref}[1]{Fig.~\ref{fig:#1}}

\newcommand{\CITE}[1]{\cite{#1}}

\begin{document}
\title[Rotational behavior of red blood cells in
suspension]{Rotational behavior of red blood cells in suspension---a
  mesoscale simulation study}

\author[F. Janoschek, F. Mancini, J. Harting, and
F. Toschi]{F. Janoschek$^1$, F. Mancini$^2$, J. Harting$^{1,3}$, and
  F. Toschi$^{1,4}$}

\affiliation{$^1$ Department of Applied Physics and J.~M.~Burgers
  Centre for Fluid Dynamics, Eindhoven University of
  Technology, P.\,O. Box 513, 5600\,MB Eindhoven, The Netherlands\\
  $^2$ Department of Physics and INFM, University of Rome ``Tor
  Vergata'',
  Via della Ricerca Scientifica 1, 00133 Rome, Italy\\
  $^3$ Institute for Computational Physics, University of Stuttgart,
  Pfaffenwaldring 27, 70569 Stuttgart, Germany\\
  $^4$ Department of Mathematics and Computer Science, Eindhoven
  University of Technology, P.\,O. Box 513, 5600\,MB Eindhoven, The
  Netherlands}
\label{firstpage}
\maketitle
\begin{abstract}{Coarse-grained hemodynamics, Lattice Boltzmann
    simulation, Viscosity measurement, Cell orientation, Statistical
    analysis, Soft particle suspension}
  The nature of blood as a suspension of red blood cells makes
  computational hemodynamics a demanding task. Our coarse-grained
  blood model, which builds on a lattice Boltzmann method for soft
  particle suspensions, enables the study of the collective behavior
  of the order of $10^6$ cells in suspension. After demonstrating the
  viscosity measurement in Kolmogorov flow, we focus on the
  statistical analysis of the cell orientation and rotation in Couette
  flow. We quantify the average inclination with respect to the flow
  and the nematic order as a function of shear rate and hematocrit. We
  further record the distribution of rotation periods around the
  vorticity direction and find a pronounced peak in the vicinity of
  the theoretical value for free model cells even though cell-cell
  interactions manifest themselves in a substantial width of the
  distribution.
\end{abstract}

\section{Introduction}

Human blood can be approximated as a suspension of deformable red
blood cells (RBCs, also called erythrocytes) in a Newtonian liquid,
the blood plasma. The other constituents like leukocytes and
thrombocytes can be neglected due to their low volume
concentrations. Typical physiological RBC concentrations are around
$50\,\%$. In the absence of external stresses, erythrocytes assume the
shape of biconcave discs of approximately $8\,\mu\mathrm{m}$ diameter
\CITE{fung81}. An understanding of their effect on the rheology and
the clotting behavior of blood is necessary for the study of
pathological deviations in the body and the design of microfluidic
devices for improved blood analysis.

Well-established methods for the computer simulation of blood flow
either consist of an elaborate model of deformable cells
\CITE{noguchi05,dupin07} or restrict themselves to a continuous
description at larger scales \CITE{boyd07}. Our motivation is to
bridge the gap between both classes of models by an intermediate
approach: we keep the particulate nature of blood but simplify the
description of each cell as far as possible. The main idea of the
model~\CITE{janoschek10} is to distinguish between the long-range
hydrodynamic coupling of cells and their complex short-range
interactions. Our method is well suited for the implementation of
complex boundary conditions and an efficient parallelization on
parallel supercomputers. This is important for the accumulation of
statistically relevant data to link bulk properties, for example the
effective viscosity, to phenomena at the level of single
erythrocytes. In this article, we present an alternative method of
viscosity measurement and a study of the orientational dynamics of
cells in a sheared suspensions. The improved understanding of the
dynamic behavior of blood might be used for the optimization of
macroscopic simulation methods.

\section{Coarse-grained model for blood flow simulations}

We apply a D3Q19/BGK lattice Boltzmann method for modeling the blood
plasma \CITE{succi01}. The single particle distribution function
$n_r(\mathbf{x},t)$ resembles the fluid traveling with one of
$r=1,\ldots,19$ discrete velocities $\mathbf{c}_r$ at the
three-dimensional lattice position $\mathbf{x}$ and discrete time
$t$. Its evolution in time is determined by the lattice Boltzmann
equation
\begin{equation}
  \label{eq:lbe}
  n_r(\mathbf{x}+\mathbf{c}_r,t+1)
  =
  n_r(\mathbf{x},t)
  -
  \Omega
  \mbox{\quad with\quad}
  \Omega
  =
  \frac
  {n_r(\mathbf{x},t)-
    n_r^\mathrm{eq}(\rho(\mathbf{x},t),\mathbf{u}(\mathbf{x},t))}
  {\tau}
\end{equation}
being the BGK-collision term with a single relaxation time $\tau$. The
equilibrium distribution function
$n_r^\mathrm{eq}(\rho,\mathbf{u})$ is an expansion of the
Maxwell--Boltzmann distribution.
$\rho(\mathbf{x},t)=\sum_rn_r(\mathbf{x},t)$ and
$\rho(\mathbf{x},t)\mathbf{u}(\mathbf{x},t)=\sum_rn_r(\mathbf{x},t)\mathbf{c}_r$
can be identified as density and momentum. In the limit of small
velocities and lattice spacings the Navier--Stokes equations are
recovered with a kinematic viscosity of $\nu=(2\tau-1)/6$, where
$\tau=1$ in this study.

For a coarse-grained description of the hydrodynamic interaction of
cells and blood plasma, a method similar to the one by Aidun et
al. modeling rigid particles of finite size is applied
\CITE{aidun98}. It can be seen as a bounce-back boundary condition
which is enhanced by a correction term
$C=2\alpha_{c_r}\rho(\mathbf{x}+\mathbf{c}_r,t)\,\mathbf{c}_r\mathbf{v}/c_\mathrm{s}^2$
that accounts for a possible local boundary velocity $\mathbf{v}$.
The lattice weights $\alpha_{c_r}$ and the speed of sound
$c_\mathrm{s}$ are constants for a given set of discrete velocities.
The resulting bounce-back rule
\begin{equation}\label{eq:lbe-bb-corr}
  n_r(\mathbf{x}+\mathbf{c}_r,t+1)
  =
  n^*_{\bar{r}}(\mathbf{x}+\mathbf{c}_r,t)
  +
  C
  \mbox{\quad with\quad}
  n^*_r(\mathbf{x},t)
  =
  n_r(\mathbf{x},t)
  -
  \Omega
\end{equation}
and $\bar{r}$ specifying the direction opposite to $r$ is applied to
all fluid distributions that according to \eqnref{lbe} would travel
along a link that crosses the theoretical cell surface. The momentum
$\Delta\mathbf{p}_\mathrm{fp}=\left(2n_{\bar{r}}+C\right)\mathbf{c}_{\bar{r}}$
which is transferred during each time step by each single bounce-back
process is used to calculate the resulting force on the boundary.

Instead of the biconcave equilibrium shape of physiological RBCs we
choose a simplified ellipsoidal geometry that is defined by two
distinct half-axes $R_\parallel$ and $R_\perp$ parallel and
perpendicular to the unit vector $\hat{\mathbf{o}}_i$ which points
along the direction of the axis of rotational symmetry of each cell
$i$. Since the resulting volumes are rigid we allow them to overlap in
order to account for the deformability of real erythrocytes. We assume
a pair of mutual forces $\mathbf{F}_\mathrm{pp}=
\pm2n_r^\mathrm{eq}(\bar{\rho},\mathbf{u}=\mathbf{0})\mathbf{c}_r$ at
each cell-cell link. This is exactly the momentum transfer during one
time step due to the rigid-particle algorithm for a resting particle
and an adjacent site with resting fluid at equilibrium and initial
density $\bar{\rho}$. While these surface forces solely serve to
compensate the lack of fluid pressure inside the particles, we
additionally implement phenomenological pair potentials for
neighboring cells that model the volume exclusion of soft and
anisotropic particles in an effective way. For the sake of simplicity,
we use the repulsive branch of a Hookian spring potential
\begin{equation}\label{eq:phi}
  \phi(r_{ij})
  =
  \left\{
    \begin{array}{l@{\qquad}l}
      \varepsilon\left(1-r_{ij}/\sigma\right)^2 &
      r_{ij}<\sigma\\
      0 &
      r_{ij}\ge\sigma
    \end{array}
  \right.
\end{equation}
for the scalar displacement $r_{ij}$ of two cells $i$ and $j$. With
respect to the disc-shape of RBCs, we follow the approach of Berne and
Pechukas \CITE{berne72} and choose the energy and range parameters
\begin{equation}\label{eq:epsilon}
  \varepsilon(\hat{\mathbf{o}}_i,\hat{\mathbf{o}}_j)
  =
  \bar{\varepsilon}
  \left[
    1-\chi^2\left(\hat{\mathbf{o}}_i\hat{\mathbf{o}}_j\right)^2
  \right]^{-1/2}
\end{equation}
and
\begin{equation}\label{eq:sigma}
  \sigma(\hat{\mathbf{o}}_i,\hat{\mathbf{o}}_j,\hat{\mathbf{r}}_{ij})
  =
  \bar{\sigma}
  \left\{
    1-\frac{\chi}{2}\left[
      \frac
      {\left(
          \hat{\mathbf{r}}_{ij}\hat{\mathbf{o}}_i
          +
          \hat{\mathbf{r}}_{ij}\hat{\mathbf{o}}_j
        \right)^2}
      {1+\chi\hat{\mathbf{o}}_i\hat{\mathbf{o}}_j}
      +
      \frac
      {\left(
          \hat{\mathbf{r}}_{ij}\hat{\mathbf{o}}_i
          -
          \hat{\mathbf{r}}_{ij}\hat{\mathbf{o}}_j
        \right)^2}
      {1-\chi\hat{\mathbf{o}}_i\hat{\mathbf{o}}_j}
    \right]
  \right\}^{-1/2}
\end{equation}
as functions of the orientations $\hat{\mathbf{o}}_i$ and
$\hat{\mathbf{o}}_j$ of the cells and their normalized center
displacement $\hat{\mathbf{r}}_{ij}$. We achieve an anisotropic
potential with a zero-energy surface that is approximately that of
ellipsoidal discs. Their half-axes parallel $\sigma_\parallel$ and
perpendicular $\sigma_\perp$ to the symmetry axis enter
\eqnref{epsilon} and \eqnref{sigma} via $\bar{\sigma}=2\sigma_\perp$
and
$\chi=(\sigma_\parallel^2-\sigma_\perp^2)/(\sigma_\parallel^2+\sigma_\perp^2)$
whereas $\bar{\varepsilon}$ determines the potential strength.

\begin{figure}
  \centering
  \includegraphics[width=0.6\columnwidth]
  {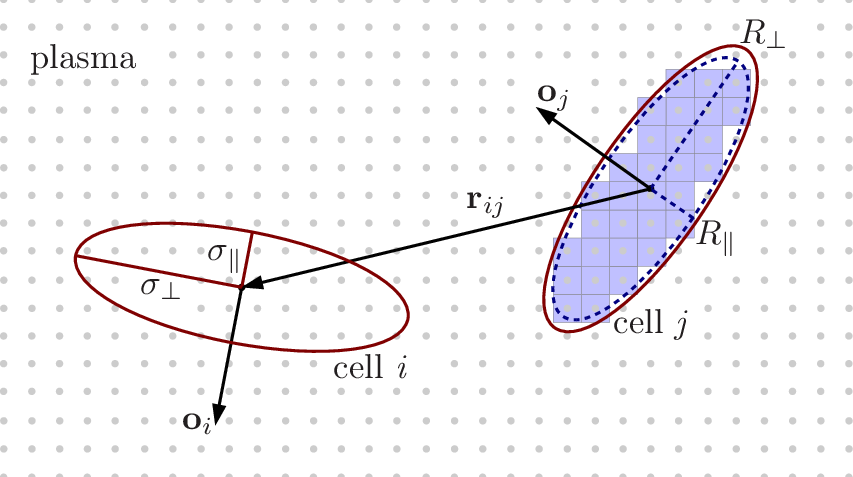}
  \caption{\label{fig:model-2d}Outline of the model. Shown are two
    cells with their axes of rotational symmetry depicted by
    vectors. The volumes defined by the cell-cell interaction are
    approximately ellipsoidal (---). The smaller ellipsoidal volumes
    (-~-~-) of the cell-plasma interaction are discretized on the
    underlying lattice. (Online version in color.)}
\end{figure}

\picref{model-2d} shows an outline of the model. Two cells $i$ and $j$
surrounded by blood plasma are displayed. For clarity, the
three-dimensional model is visualized in a two-dimensional
cut. Depicted are the cell shapes defined by the zero-energy surface
of the cell-cell potential \eqnref{phi} with \eqnref{epsilon} and
\eqnref{sigma} that can be approximated by ellipsoids with the size
parameters $\sigma_\parallel$ and $\sigma_\perp$ as half axes. The
cells are free to assume continuous positions and orientations
$\mathbf{o}_i$ and $\mathbf{o}_j$. In consequence, also the center
displacement vectors $\mathbf{r}_{ij}$ and $\mathbf{r}_{i\mathbf{x}}$
between cells are continuous. Still, for the cell-plasma interaction
an ellipsoidal volume with half axes $R_\parallel$ and $R_\perp$ is
discretized on the underlying lattice. For additional clarity, the
discretization is drawn only for cell $j$. The forces and torques
emerging from the interaction of the cells with other RBCs and the
fluid are integrated by a classical molecular dynamics code in order
to evolve the system in time. The conversion from lattice units to
physical units is done by multiplication with $\delta
x=2/3\,\mu\mathrm{m}$, $\delta t=6.80\times10^{-8}\,\text{s}$, and
$\delta m=3.05\times10^{-16}\,\text{kg}$ for length, time, and mass
respectively. As a convention, primed variables are used whenever we
refer to quantities specified in physical units. The model parameters
are chosen as $R'_\perp=11/3\,\mu\text{m}$,
$R'_\parallel=11/9\,\mu\text{m}$, $\sigma'_\perp=4\,\mu\text{m}$,
$\sigma'_\parallel=4/3\,\mu\text{m}$, and
$\bar{\varepsilon}'=1.47\times10^{-15}\,\mathrm{J}$. For more detailed
information see~\CITE{janoschek10}.

\section{Results}

\subsection{Apparent bulk viscosity measurement in Kolmogorov flow}
In our previous work, the apparent viscosity of the bulk of the
suspension was measured in simulations of unbounded Couette
flow~\CITE{janoschek10}. Though this setting is easy to comprise
analytically, its efficient and precise implementation proves a
non-trivial challenge since the suspended particles would have to be
enabled to stretch across the sheared boundary. We therefore
demonstrate an alternative method to determine the apparent viscosity
that is compatible with completely periodic boundaries and is based on
so-called \textit{Kolmogorov flow}: a sinusoidally modulated shear
flow. We proceed as in Benzi \textit{et al.}~\CITE{benzi10} and apply
to the full suspension a sinusoidal body force
\begin{equation}\label{eq:fz}
  f_z(x)=f_0\sin\left[k(x-1/2)\right]
\end{equation}
pointing into the $z$-direction. $f_0$ is the amplitude and $k$ the
wave number in $x$-direction. At steady state, the spatial variation
of the shear stress
\begin{equation}\label{eq:dsigma}
  \partial_x\sigma_{xz}(x)=f_z(x)
\end{equation}
matches the external forcing. Integration of \eqnref{dsigma} together
with \eqnref{fz} yields an analytic expression for the shear stress
\begin{equation}
  \sigma_{xz}(x)=-\frac{f_0}{k}\cos\left[k(x-1/2)\right]
  \text{ .}
\end{equation}
The $y$- and $z$-averaged local shear rate can be evaluated
numerically as
\begin{equation}\label{eq:gammadot}
  \dot{\gamma}_{xz}(x)=\left\langle\partial_xu_z(x)\right\rangle_{y,z}\text{ .}
\end{equation}
After equilibration, a simulation of $N_x\times N_y\times N_z$ lattice
sites results in $N_x$ numbers for the apparent viscosity
\begin{equation}
  \mu_\text{app}(x)=\frac{\sigma_{xz}(x)}{\dot{\gamma}_{xz}(x)}
\end{equation}
covering a range of shear rates that depends on $f_0$ and $k$.

Compared to a viscosity measurement in Couette
flow~\CITE{janoschek10}, the above procedure~\CITE{benzi10} has the
benefit of taking advantage of the whole simulation volume since there
are no possibly unphysical boundary regions. Additionally, each single
measurement yields data for many different shear rates. On the other
side, the non-constant shear stress causes inhomogeneities in the
local cell volume concentration $\Phi(x)$ which are much more
pronounced than in the case of Couette flow. Therefore, we disregard
all viscosity data $(x, \mu_\text{app}(x))$ for which
$\Phi(x)\not\in[\Phi^*-\Delta\Phi,\Phi^*+\Delta\Phi]$ with $\Phi^*$
being the volume concentration of interest and $2\Delta\Phi$ a small
interval of tolerance. \picref{kolmogorov} compares
$\mu_\text{app}(|\dot{\gamma}_{xz}|)$ resulting from such measurements
at varying $f_0$ and $N_x$ with $k=2\pi/N_x$ after $68.0\,\mathrm{ms}$
with Couette flow measurements as described in \CITE{janoschek10}
($N_x=128$). In the plot, $\Phi^*=43\,\%$ and $\Delta\Phi=0.5\,\%$. We
further average the data within bins with a width of
$\Delta[\ln(\dot{\gamma}'_{xz}\,\mathrm{s})]=0.5$ in order to reduce
statistical noise. The good agreement proves the feasibility of the
method.

\begin{figure}
  \centering
  \includegraphics[width=\columnwidth]{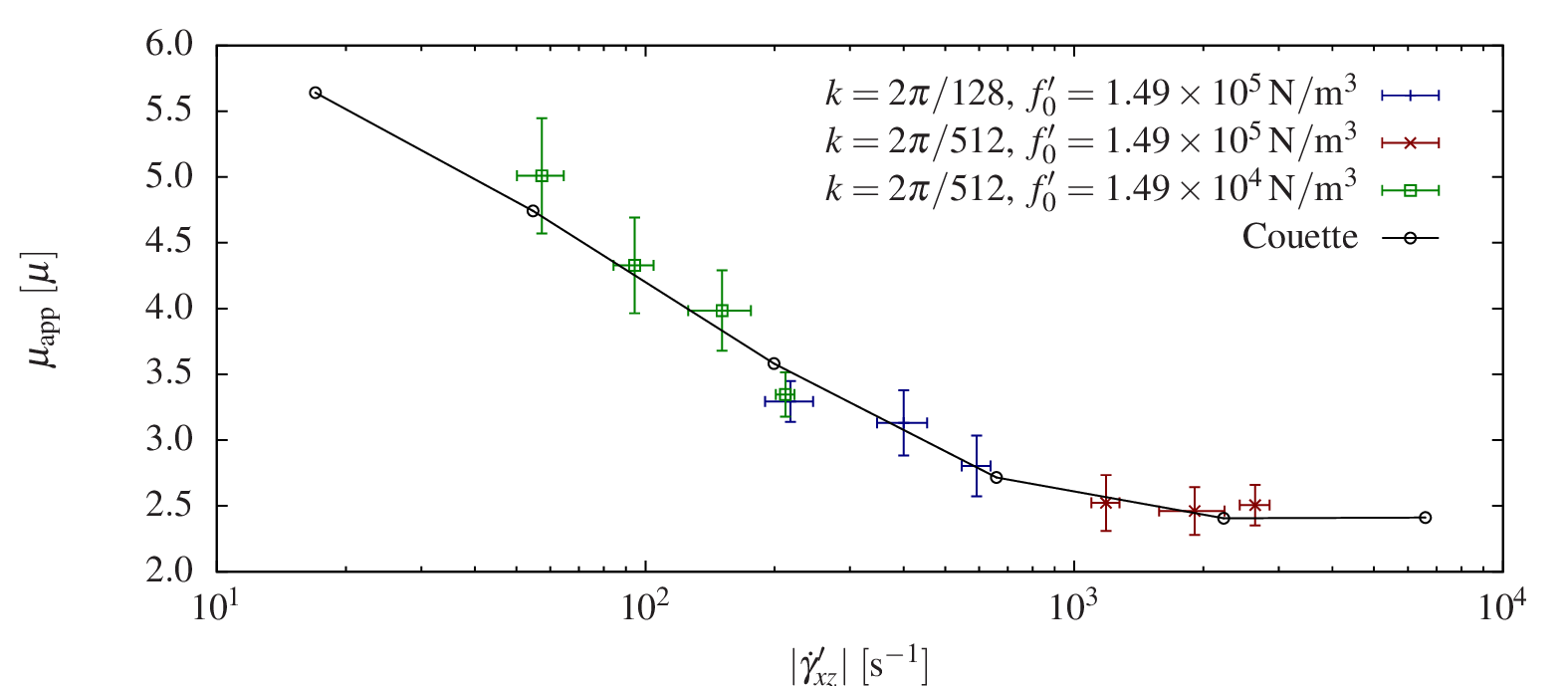}
  \caption{\label{fig:kolmogorov}Comparison of viscosity data
    retrieved from Kolmogorov flow at different wave numbers $k$ and
    amplitudes of forcing $f'_0$ after $68.0\,\mathrm{ms}$ of
    equilibration with data measured in Couette flow as
    in~\CITE{janoschek10}. The hematocrit is $\Phi^*=43\,\%$ with
    $\Delta\Phi=0.5\,\%$ while the error bars were drawn according to
    the standard deviation after binning and averaging of the original
    data. (Online version in color.)}
\end{figure}

\subsection{Statistical analysis of cell orientations in Couette flow}

Due to its applicability to huge numbers of particles and the fact
that cell orientations are accessible directly in the model, our
simulation method is particularly suited for the statistical analysis
of RBC orientations at high volume concentrations. In the following,
we simulate Couette flow as described in~\CITE{janoschek10}. However,
with a size of $N_x=N_z=512$ and $N_y=64$ lattice sites, the volume is
considerably larger and contains about $3\times10^4$ cells at a
hematocrit of $\Phi=45\,\%$. The velocity gradient is aligned with
$x$- and the flow with the $z$-axis. To exclude boundary effects in
the $x$ direction, we restrict ourselves to the analysis of those
cells which stay at a lateral position $512/3<x<2\times512/3$ during
the whole simulation. The visualization of the cell volumes defined by
the model potential is shown for a smaller system in the inset of
\picref{orientation}. It is clear that also at physiological volume
concentrations there is a preferential orientation of cells. In order
to quantify this observation, the orientation vector of every cell $i$
is multiplied with $-1$ where necessary to achieve
$(\hat{\mathbf{o}}_i)_x>0$. The angle $\theta$ is measured between the
positive $x$-axis and the resulting normalized orientation
vector. \picref{orientation} displays the most probable value
$\theta^*$ that is determined by fitting a Gaussian to the
distribution of angles. For $\Phi=45\,\%$, compared to $\theta=0$
which would mean alignment parallel to the flow, there always is a
positive inclination $\theta^*$ which is directed against the
vorticity of the shear (see inset of \picref{orientation}). When
increasing the shear rate from around $300\,\mathrm{s}^{-1}$ to
$6000\,\mathrm{s}^{-1}$, this inclination becomes considerably
smaller: $\theta^*$ decreases from $14.5\!\degree$ to $3.4\!\degree$.

\begin{figure}
  \centering
  \includegraphics[width=\columnwidth]{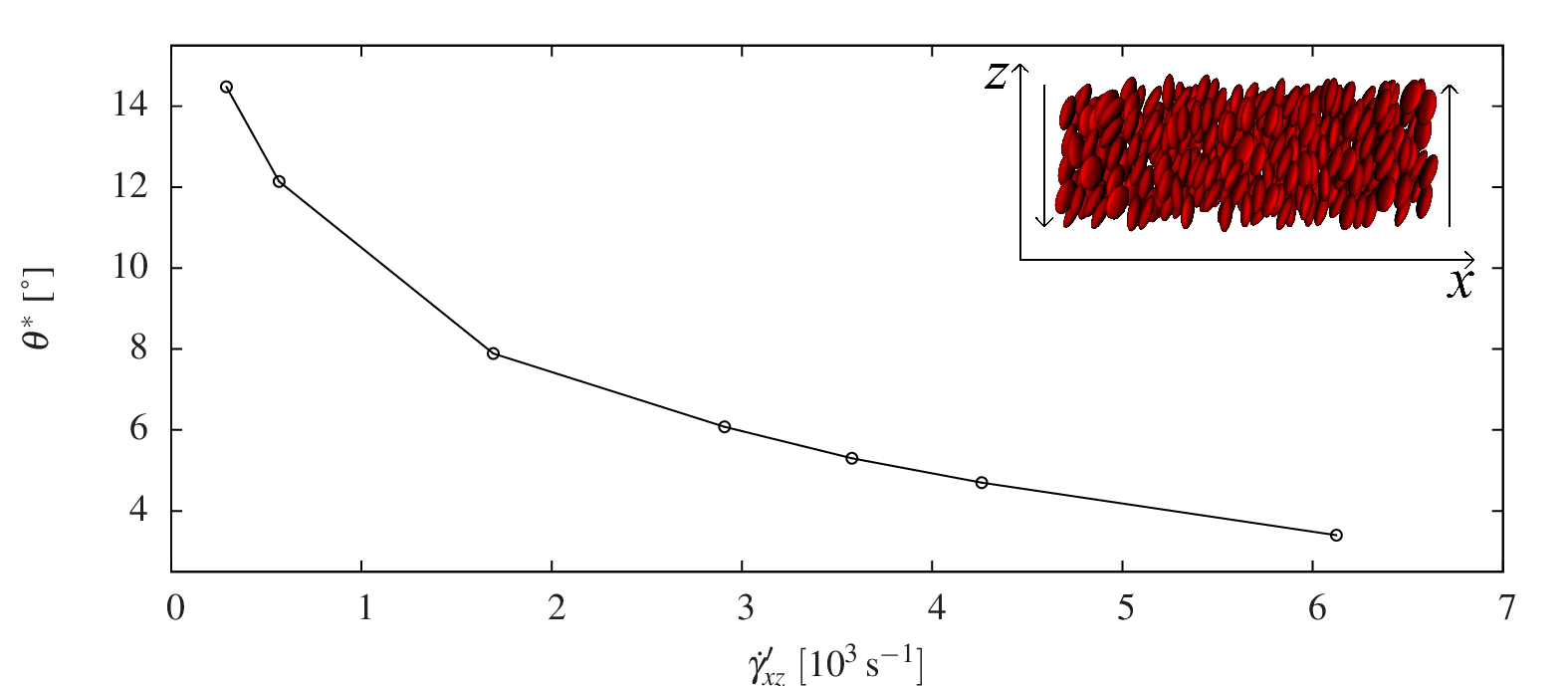}
  \caption{\label{fig:orientation}Most probable value of the angle
    $\theta$ between the axis of symmetry of a model cell and the
    $x$-direction as a function of the shear rate
    $|\dot{\gamma}'_{xz}|$ at $\Phi=45\,\%$. The statistical error of
    $\theta$ is smaller than the symbols. The inset shows alignment of
    cells due to shear flow in a smaller system. (Online version in
    color.)}
\end{figure}

Studying $\theta$ allows to quantify the orientation with respect to
the direction of the flow. To quantify the actual amount of
orientational order, the nematic order parameter known from liquid
crystal physics (see for example~\CITE{tsige99}) is a better
choice. This parameter is usually defined as the largest eigenvalue
$\lambda_+$ of the nematic order tensor
$S_{kl}=\left\langle3\hat{o}_k\hat{o}_l-\delta_{kl}\right\rangle_{i,t}/2$
which is obtained as the average over different time steps $t$ and
cells $i$ within the volume of interest. Possible values for
$\lambda_+$ are comprised between 0 and 1 indicating complete disorder
and order. \picref{nematic} depicts $\lambda_+$ for different shear
rates. A decrease of nematic order with increasing shear rate is
visible. The inset shows the dependency on the hematocrit $\Phi$ at a
fixed shear rate of
$\dot{\gamma}'_{xz}=(3.0\pm0.1)\times10^3\,\mathrm{s}^{-1}$. For
comparison, also $\lambda_+$ resulting for a single ellipsoidal
particle that tumbles with $\hat{\mathbf{o}}$ perpendicular to the
vorticity direction is shown. Apparently, the higher nematic order at
$\Phi=45\,\%$ is caused by hindered tumbling motion while the reason
for the lower $\lambda_+$ at $\Phi=11\,\%$ is the lack of alignment in
the vorticity plane.

\begin{figure}
  \centering
  \includegraphics[width=\columnwidth]{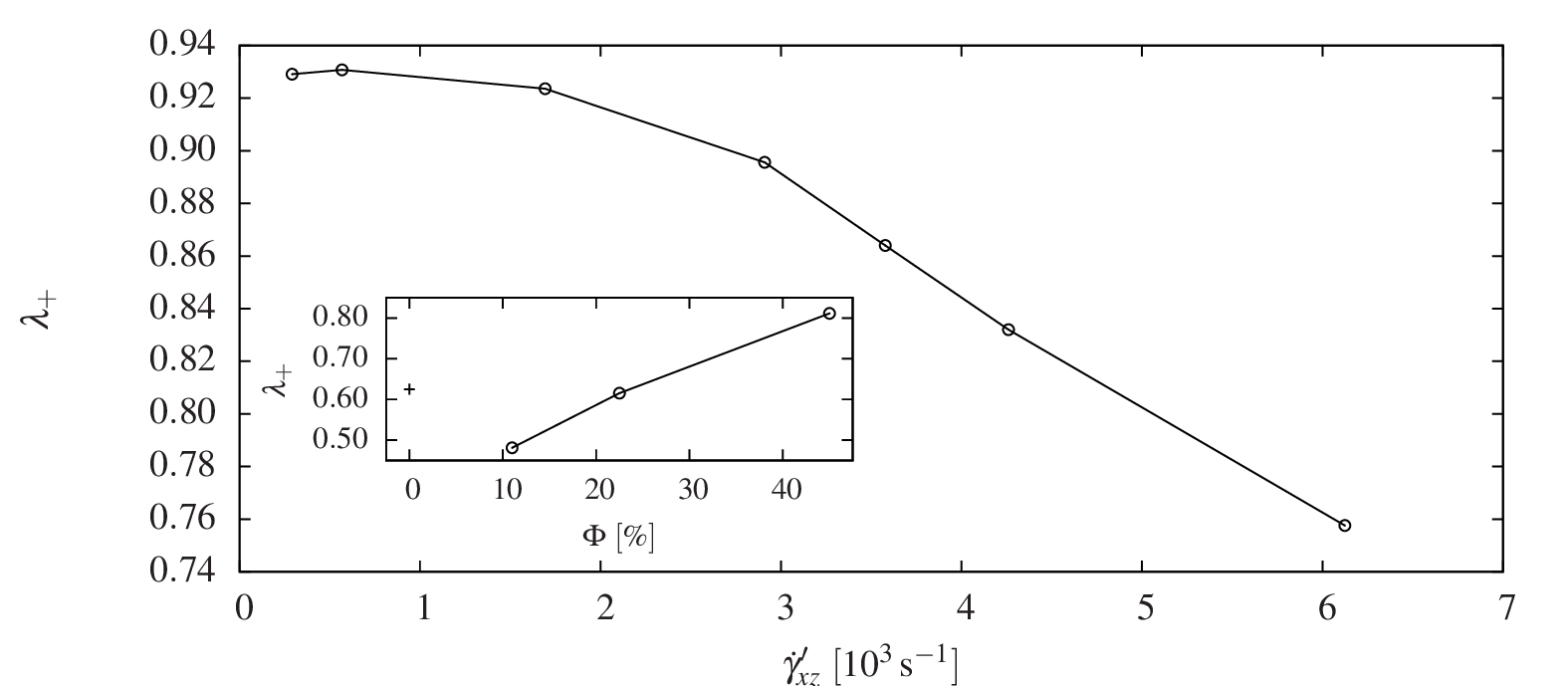}
  \caption{\label{fig:nematic}Nematic order parameter $\lambda_+$ as a
    function of the shear rate $\dot{\gamma}'_{xz}$ at
    $\Phi=45\,\%$. In the inset,
    $\dot{\gamma}'_{xz}=(3.0\pm0.1)\times10^3\,\mathrm{s}^{-1}$ is
    kept fixed and the hematocrit $\Phi$ is varied. The single data
    point at $\Phi\to0$ is calculated for a single ellipsoidal
    particle tumbling exactly within the $xz$-plane~\CITE{jeffery22}.}
\end{figure}

\subsection{Statistical analysis of cell rotations in Couette flow}

The above conclusions make clear that preferential orientations result
from the averaged tumbling of many cells. The time evolution of the
continuous tumbling angle $\theta$ is plotted for an arbitrary
selection of cells in the inset of \picref{tumbling}. The respective
volume concentration is $\Phi=45\,\%$ and the shear rate
$\dot{\gamma}'_{xz}=6.1\times10^3\,\mathrm{s}^{-1}$. The cells keep a
largely constant alignment for varying periods of time and,
occasionally, flip over by an angle of $\pi$ along the vorticity
direction. Thus, the angular velocity is strongly time-dependent. The
probability distribution function of the time required for a rotation
by $\pi$ which is measured as the time $T$ between two flipping events
is shown in the main part of \picref{tumbling}. For decreasing volume
concentrations, the distribution becomes narrower and its peak is
shifted towards shorter times. However, with a width comparable to its
average value, even at $\Phi=11\,\%$ the distribution deviates
substantially from a delta peak. Therefore, even though typical
tumbling periods in suspension do not differ much from the value for a
freely tumbling ellipsoid~\CITE{jeffery22}, the effect of cell-cell
interactions on our observables is significant for all $\Phi$ studied.

\begin{figure}
  \centering
  \includegraphics[width=\columnwidth]{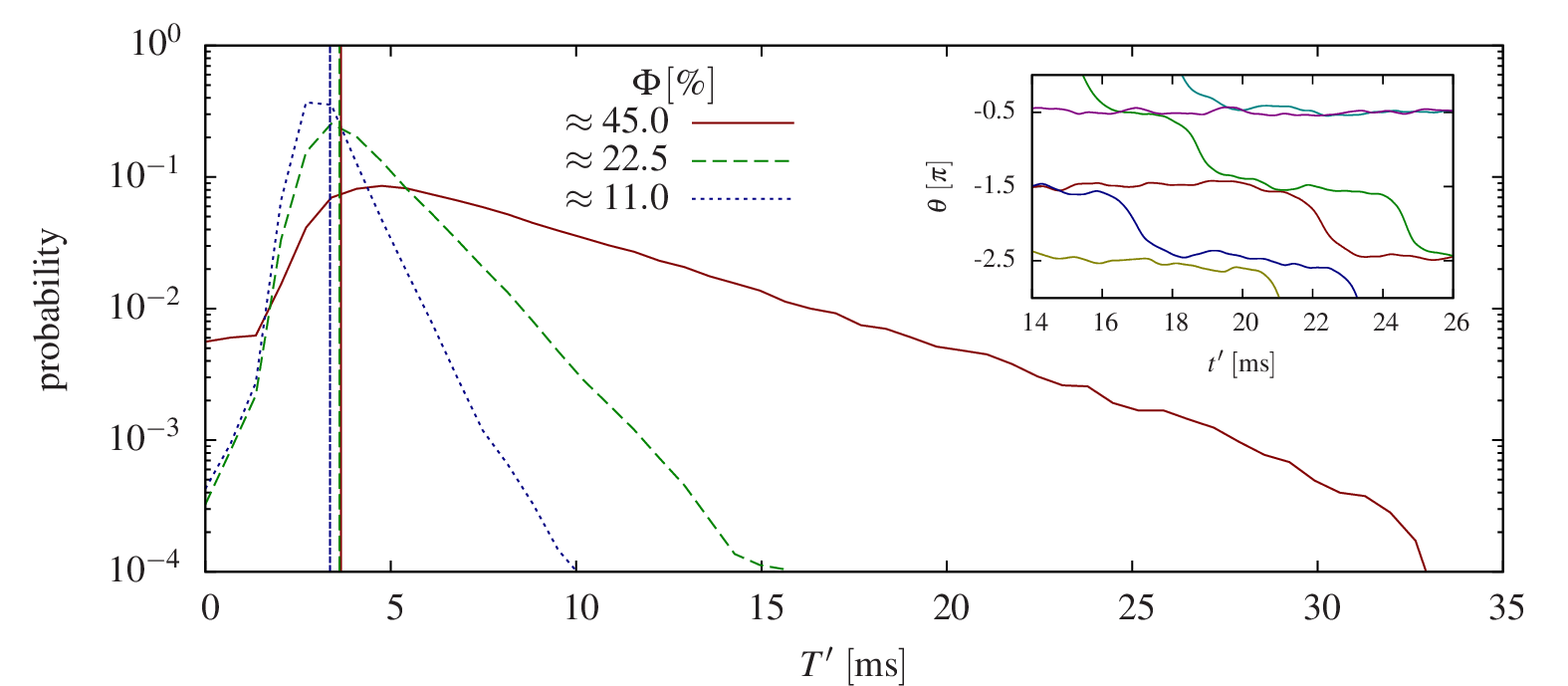}
  \caption{\label{fig:tumbling}Probability density functions of the
    time $T'$ between two consecutive events of flipping by an angle
    of $\pi$ around the vorticity direction at different values of
    $\Phi$ and
    $\dot{\gamma}'_{xz}=(3.0\pm0.1)\times10^3\,\mathrm{s}^{-1}$. As a
    reference, the vertical lines indicate the time required for a
    rotation around $\pi$ according to the analytical solution for a
    single ellipsoidal particle at the respective shear
    rates~\CITE{jeffery22}. The inset visualizes the continuous
    tumbling angle $\theta$ of selected cells as a function of time
    $t'$ at $\Phi=45\,\%$. (Online version in color.)}
\end{figure}

\section{Conclusion}

Our mesoscale model for blood~\CITE{janoschek10} was applied to bulk
flow of up to $3\times10^4$ cells. We demonstrated the feasibility of
viscosity measurement in Kolmogorov flow and studied the statistics of
the rotational cell positions and velocities at different values of
the hematocrit. These observables are extendable also to more
elaborate (e.\,g. deformable) cell models. Their quantitative study
and measurement is a key to build more reliable cell based or
continuum descriptions for blood.

\begin{acknowledgements}
  The authors acknowledge financial support from the Netherlands
  Organization for Scientific Research (VIDI grant of
  J.~Harting) 
  and the TU/e High Potential Research Program as well as computing
  resources from JSC J\"ulich, SSC Karlsruhe, CSC Espoo, EPCC
  Edinburgh, and SARA Amsterdam, the latter three being granted by
  DEISA as part of the DECI-5 project.
\end{acknowledgements}


\begin{thebibliography}{15}
\providecommand{\natexlab}[1]{#1}
\expandafter\ifx\csname urlstyle\endcsname\relax
  \providecommand{\doi}[1]{DOI \discretionary{}{}{}#1}\else
  \providecommand{\doi}{DOI \discretionary{}{}{}\begingroup
  \urlstyle{rm}\Url}\fi
\bibitem{fung81}
Fung, Y.~C. 1981 \emph{Biomechanics. {M}echanical properties of living
  tissues}.
\newblock New York: Springer, 1st edn.

\bibitem{noguchi05}
Noguchi, H. \& Gompper, G. 2005 Shape transitions of fluid vesicles and red
  blood cells in capillary flows.
\newblock \emph{PNAS}, \textbf{102}, 14\,159--14\,164.
\newblock (DOI 10.1073/pnas.0504243102)

\bibitem{dupin07}
Dupin, M.~M., Halliday, I., Care, C.~M., Alboul, L. \& Munn, L.~L. 2007
  Modeling the flow of dense suspensions of deformable particles in three
  dimensions.
\newblock \emph{Phys. Rev. E}, \textbf{75}, 066\,707.
\newblock (DOI 10.1103/PhysRevE.75.066707)

\bibitem{boyd07}
Boyd, J., Buick, J.~M. \& Green, S. 2007 Analysis of the {C}asson and
  {C}arreau-{Y}asuda non-{N}ewtonian blood models in steady and oscillatory
  flows using the lattice {B}oltzmann method.
\newblock \emph{Phys. Fluids}, \textbf{19}, 093\,103.
\newblock (DOI 10.1063/1.2772250)

\bibitem{janoschek10}
Janoschek, F., Toschi, F. \& Harting, J. 2010 A simplified particulate model
  for coarse-grained hemodynamics simulations.
\newblock See http://arxiv.org/abs/1005.2594v2.

\bibitem{succi01}
Succi, S. 2001 \emph{The lattice {B}oltzmann equation for fluid dynamics and
  beyond}.
\newblock Oxford University Press, 1st edn.

\bibitem{aidun98}
Aidun, C.~K., Lu, Y. \& Ding, E.-J. 1998 Direct analysis of particulate
  suspensions with inertia using the discrete {B}oltzmann equation.
\newblock \emph{J. Fluid Mech.}, \textbf{373}, 287--311.

\bibitem{berne72}
Berne, B.~J. \& Pechukas, P. 1972 Gaussian model potentials for molecular
  interactions.
\newblock \emph{J. Chem. Phys.}, \textbf{56}, 4213--4216.
\newblock (DOI 10.1063/1.1677837)

\bibitem{benzi10}
Benzi, R., Bernaschi, M., Sbragaglia, M. \& Succi, S. 2010 {Herschel-Bulkley}
  rheology from lattice kinetic theory of soft-glassy materials.
\newblock See http://arxiv.org/abs/1004.5058v1.

\bibitem{tsige99}
Tsige, M., Mahajan, M.~P., Rosenblatt, C. \& Taylor, P.~L. 1999 Nematic order
  in nanoscopic liquid crystal droplets.
\newblock \emph{Phys. Rev. E}, \textbf{60}, 638--644.
\newblock (DOI 10.1103/PhysRevE.60.638)

\bibitem{jeffery22}
Jeffery, G.~B. 1922 The motion of ellipsoidal particles immersed in a viscous
  fluid.
\newblock \emph{Proc. R. Soc. Lond. A}, \textbf{102}, 161--179.
\end{thebibliography}

\label{lastpage}
\end{document}